\documentclass{emulateapj}
\slugcomment{To be publichsed in ApJ main journal.}

\shorttitle{TeV Gamma Rays from PSR~B1706$-$44} 
\shortauthors{R. Enomoto et al.}

\begin{document}

\title{CANGAROO-III Observation of TeV Gamma Rays from the vicinity of PSR~B1706$-$44}

\author{
R.~Enomoto\altaffilmark{1}
J.~Kushida\altaffilmark{2}
T.~Nakamori\altaffilmark{3}
T.~Kifune\altaffilmark{1}
G.~V.~Bicknell\altaffilmark{4}
R.~W.~Clay\altaffilmark{5}
P.~G.~Edwards\altaffilmark{6}
S.~Gunji\altaffilmark{7}
S.~Hara\altaffilmark{8}
T.~Hara\altaffilmark{9}
T.~Hattori\altaffilmark{2}
S.~Hayashi\altaffilmark{10}
Y.~Higashi\altaffilmark{11}
Y.~Hirai\altaffilmark{12}
K.~Inoue\altaffilmark{7}
H.~Ishioka\altaffilmark{2}
S.~Kabuki\altaffilmark{11}
F.~Kajino\altaffilmark{10}
H.~Katagiri\altaffilmark{13}
A.~Kawachi\altaffilmark{2}
R.~Kiuchi\altaffilmark{1}
H.~Kubo\altaffilmark{11}
T.~Kunisawa\altaffilmark{1}
T.~Matoba\altaffilmark{12}
Y.~Matsubara\altaffilmark{14}
I.~Matsuzawa\altaffilmark{2}
T.~Mizukami\altaffilmark{11}
Y.~Mizumura\altaffilmark{2}
Y.~Mizumoto\altaffilmark{15}
M.~Mori\altaffilmark{1}
H.~Muraishi\altaffilmark{16}
T.~Naito\altaffilmark{9}
S.~Nakano\altaffilmark{11}
K.~Nishijima\altaffilmark{2}
M.~Ohishi\altaffilmark{1}
Y.~Otake\altaffilmark{7}
S.~Ryoki\altaffilmark{1}
K.~Saito\altaffilmark{2}
Y.~Sakamoto\altaffilmark{2}
A.~Seki\altaffilmark{2}
V.~Stamatescu\altaffilmark{5}
T.~Suzuki\altaffilmark{12}
D.~L.~Swaby\altaffilmark{5}
T.~Tanimori\altaffilmark{11}
G.~Thornton\altaffilmark{5}
F.~Tokanai\altaffilmark{7}
K.~Tsuchiya\altaffilmark{17}
S.~Watanabe\altaffilmark{11}
E.~Yamazaki\altaffilmark{2}
S.~Yanagita\altaffilmark{12}
T.~Yoshida\altaffilmark{12}
T.~Yoshikoshi\altaffilmark{1}
Y.~Yukawa\altaffilmark{1}
}

\altaffiltext{1}{ Institute for Cosmic Ray Research, University of Tokyo, Kashiwa, Chiba 277-8582, Japan} 
\altaffiltext{2}{ Department of Physics, Tokai University, Hiratsuka, Kanagawa 259-1292, Japan} 
\altaffiltext{3}{ Department of Basic Physics, Tokyo Institute of Technology, Meguro, Tokyo, 152-8551, Japan} 
\altaffiltext{4}{ Research School of Astronomy and Astrophysics, Australian National University, ACT 2611, Australia} 
\altaffiltext{5}{ School of Chemistry and Physics, University of Adelaide, SA 5005, Australia} 
\altaffiltext{6}{ Narrabri Observatory of the Australia Telescope National Facility, CSIRO, Epping, NSW 2121, Australia} 
\altaffiltext{7}{ Department of Physics, Yamagata University, Yamagata, Yamagata 990-8560, Japan} 
\altaffiltext{8}{ Ibaraki Prefectural University of Health Sciences, Ami, Ibaraki 300-0394, Japan} 
\altaffiltext{9}{ Faculty of Management Information, Yamanashi Gakuin University, Kofu, Yamanashi 400-8575, Japan} 
\altaffiltext{10}{ Department of Physics, Konan University, Kobe, Hyogo 658-8501, Japan} 
\altaffiltext{11}{ Department of Physics, Kyoto University, Sakyo-ku, Kyoto 606-8502, Japan} 
\altaffiltext{12}{ Faculty of Science, Ibaraki University, Mito, Ibaraki 310-8512, Japan} 
\altaffiltext{13}{ Department of Physical Science, Hiroshima University, Higashi-Hiroshima, Hiroshima 739-8526, Japan} 
\altaffiltext{14}{ Solar-Terrestrial Environment Laboratory,  Nagoya University, Nagoya, Aichi 464-8602, Japan} 
\altaffiltext{15}{ National Astronomical Observatory of Japan, Mitaka, Tokyo 181-8588, Japan} 
\altaffiltext{16}{ School of Allied Health Sciences, Kitasato University, Sagamihara, Kanagawa 228-8555, Japan} 
\altaffiltext{17}{ National Research Institute of Police Science, Kashiwa, Chiba 277-0882, Japan}

\begin{abstract}

Observation   
by the CANGAROO-III stereoscopic system of the Imaging Cherenkov Telescope 
has detected extended emission of TeV gamma rays   
in the vicinity of the pulsar PSR~B1706$-$44.  
The strength of the signal observed as gamma-ray-like events varies      
when we apply different ways of emulating background events.     
%different way of analysis in 
% selecting $\lq\lq$off-source data" for estimating the background.
%when we adopt a different way of analysis in 
% selecting $\lq\lq$off-source data" for estimating the background.    
%The reason for such an ambiguity concerning
% the credibility of detection  
The reason for such uncertainties is argued 
in relevance to gamma-rays embedded in 
the $\lq\lq$off-source data", that is,  
unknown sources and diffuse emission in the Galactic plane, 
namely, the existence of 
a complex structure of TeV gamma-ray emission 
around PSR~B1706$-$44.     
\end{abstract}

\date{\today}

\section{Introduction}

PSR B1706$-$44 is a 
young pulsar with a high spin-down luminosity, 
one of the gamma-ray pulsars 
detected by the EGRET instrument of the Compton Gamma Ray Observatory
\citep{thompson}. 
Detection of point-like gamma-ray emission 
of TeV gamma-rays was reported,   
using data of the CANGAROO-I telescope,    
 at the position of 
the pulsar with a flux of $\sim 10^{-11}$ cm$^{-2}$ s$^{-1}$ 
for gamma-ray energy $>$1TeV \citep{kifune}, 
which was not pulsed in modulation with the pulsar period, and 
is thus likely, if the detection is valid, 
to be from a compact pulsar wind nebula associated 
 with PSR~B1706$-$44,  similar to the case of the Crab nebula.
This result was 
followed by the Durham Mark 6 telescope, which detected a flux of 
$\sim 4 \times 10^{-11}$ cm$^{-2}$ s$^{-1}$ at $E>300$GeV \citep{durham}.  
Observations with the CANGAROO-II telescope preliminarily reported  
the detection of a gamma-ray signal, which appeared to be somewhat broader 
than the point-spread function, $\sim 0.2^o$ of the telescope \citep{kushida}. 
 
In order to reduce background events and  
to improve the accuracy of the arrival direction, 
stereoscopic observations with 
a system of multiple Imaging Air Cherenkov Telescopes (IACTs) 
have now been put into operation as ``second 
generation IACTs", such as H.E.S.S. and VERITAS, as well as CANGAROO-III,
and 
in addition, the MAGIC telescope of a large aperture of 17m diameter.   
The sensitivity to TeV gamma rays has been dramatically improved 
to uncover a number of TeV gamma ray-sources, more than 70  
at the time of the 21st International 
Cosmic Ray Conference in 2007 \cite{hinton-rapporteur}. 

However,  
H.E.S.S. reported a null detection of
point-like emission from the vicinity of PSR B1706$-$44, 
setting an upper limit of 
$1.4 \times 10^{-12}$ cm$^{-2}$ s$^{-1}$ for $E>350$GeV 
 \citep{hess} at the center of the field of view. 
CANGAROO-III failed to confirm the signal, and obtained 
an upper limit of $5 \times 10^{-12}$ cm$^{-2}$ s$^{-1}$ for $E>600$GeV 
on the point-like emission   
 \citep{tanimori}. 
 
A number of Galactic sources of TeV gamma rays have been discovered      
through a scan survey of the Galactic disk by the H.E.S.S. Group 
\citep{hess_science}, and   
it is considered that pulsar wind nebulae (PWN) constitute 
one major class of the sources.   
A considerable number of those sources are found to exhibit extended  
emission, with its central position displaced  
 from the pulsar position in some cases.  
 It may also be the case for PSR~B1706$-$44 that  
the pulsar is accompanied by a PWN of extended TeV gamma-rays,     
or that 
some unknown object of TeV gamma-ray emission  exists in its vicinity 
without having any clear counterpart 
in other wavelengths, such as X-rays. 
 
We still have a long way to go 
before the TeV gamma-ray signal from IACT will mature to become   
completely free from the background of cosmic ray particles,    
of which intensity outnumbers TeV gamma-rays by orders of magnitude.      
In order to reveal gamma-ray events as a prominent signal,   
it is required to subtract  
background events from the observed data 
and, for this purpose, to deduce the background events  
from a dataset which is considered not to contain gamma-ray events.    

The H.E.S.S. scan survey of the Galactic disk has 
shown a broad, diffuse emission in the Galactic plane  
near the Galactic Center, when integrated over a 
solid angle of $10^{-3}$ sr,  
having an intensity as large as the flux 
from the Crab pulsar nebula         
with its spectrum as hard as $\sim E^{-2.3}$ \citep{hess_gp}. 
This fact warns us that TeV gamma-ray emission may exist 
 where no prominent objects are seen, and 
we have to be careful when we use events from an area in the Galactic disk 
to estimate the background.   

Various ways for inferring the background events 
as well as for monitoring the variance of the observation conditions  
are used in different observations.   
For example, 
the observations of CANGAROO-I, -II, and -III 
consist of two kinds of separate observations of ON-source and 
OFF-source runs. 
The OFF-source observation, 
of which direction is selected to be more than 
30$^o$ away from the source position, 
is utilized as the dataset for emulating the background in ON-source data. 
On the other hand, 
%an observation by 
the H.E.S.S. group adopts 
the method of so-called $\lq\lq$wobble" and/or $\lq\lq$ring background", 
in which    
the datasets for emulating background events are taken 
from $\lq\lq$off-source" directions of $\sim 0.5^o$ away 
from the source position   
within the common field of view of the $\lq\lq$ON-source observation".      

Presented in this paper is the result of 
CANGAROO-III observations of PSR~B1706$-$44, 
obtained through analysis 
by utilizing two different methods 
for estimating the cosmic-ray background by  
(a) the $\lq\lq$OFF-source observation",  
which is directed away from the Galactic plane,   
and (b) the $\lq\lq$wobble" and the $\lq\lq$ring background" method,  
which uses, for emulating background,  
the events coming from the directions along a ring-shaped region 
within the field of view of the ON-source run 
that contains the observation target near its center.  

\section{CANGAROO-III Stereoscopic System}

The CANGAROO-III stereoscopic system consists of four IACTs 
located near Woomera, South Australia (31$^\circ$S,
137$^\circ$E).
Each telescope has a 10-m diameter of a segmented reflector, 
consisting of 114 spherical mirrors 
made of FRP, {\it i.e.,} fiber-reinforced plastic material \citep{kawachi}, 
each of 80\,cm diameter,
mounted on a parabolic
frame with a focal length of 8\,m.
The total area for light collection is 57.3\,m$^2$.
The first telescope, T1, which has been operated as the CANGAROO-II telescope
\citep{enomoto_nature},
is not presently in use 
because of its smaller field of view than the others and
 a deterioration of the reflectivity of the plastic mirrors. 
The second, third, and fourth telescopes (T2, T3, and T4) were used for the
observations reported in this paper.
The camera and electronics system for T2, T3, and T4 and other details
are given in \citet{kabuki}.
The telescopes are located at the 
east (T1), west (T2), south (T3) and north (T4)
corners of a diamond shape 
with its sides of $\sim$100\,m \citep{enomoto_app}.

\section{Observations}

The observations of PSR~B1706$-$44 were carried out 
during the period between Jul 11 and 19 in 2004 
and between Apr 14 and Jun 15 in 2007.
During the ON-source runs in 2004, 
the center of the field of view of the telescope 
was set  
at right ascension $\alpha$=257.4$^o$ and 
declination $\delta=-44.5^o$ [J2000], 
{\it i.e.,} the position of PSR~B1706$-$44.  
In 2007,
the $\lq\lq$wobble mode" \citep{wobble} was adopted, {\it i.e.,} 
the pointing position of each telescope was displaced  
from PSR~B1706$-$44 
in declination by $\pm 0.5^o$, and was changed 
every 20 minutes to be set alternatively at $+0.5^o$ or  
 at $-0.5^o$ from $\delta=-44.5^o$. 
It is noted that we have a merit, in the wobble mode observation,  
to enlarge the effective field of view of the telescope 
and to average the response  
of the photomultiplier camera  
which fluctuates from pixel to pixel.

Listed in Table \ref{posoff} are the directions pointed by the telescope  
during the OFF-source runs as well as the date when each run was 
conducted.   
\begin{table}
\caption{Pointing positions of OFF source runs. ''Offset"
is the declination offset from PSR B1706-44. 
The direction of the center of field of the view of the telescope 
is shown in the equatorial coordinate of R.A. and $\delta$, 
as well as in Galactic latitude $b$ and longitude $l$.}
\label{posoff}
\begin{tabular}{cccccc}
\hline\hline
Date&Offset&R.A.& $\delta$ & $l$ & $b$ \\
YYYY/MM/DD & degree & degree & degree & degree & degree \\
\hline
2004/Jul/11&0& 243.89& -44.48& 336.75& 4.57\\
2004/Jul/13&0& 311.14& -44.48& 356.17& -38.35\\
2004/Jul/14&0& 334.43& -44.48& 353.19& -54.79\\
2004/Jul/17&0& 334.13& -44.48& 353.29& -54.58\\
2004/Jul/18&0& 331.53& -44.48& 354.10& -52.79\\
2004/Jul/19&0& 334.97& -44.48& 353.00& -55.16\\
2007/Apr/14&+0.5& 227.69& -43.98& 327.79& 12.06\\
2007/Apr/14&-0.5& 227.69& -44.98& 327.26& 11.20\\
2007/Apr/15&-0.5& 213.00& -44.98& 317.59& 15.58\\
2007/Apr/15&+0.5& 213.00& -43.98& 317.92& 16.53\\
2007/Apr/16&-0.5& 212.30& -44.98& 317.10& 15.73\\
2007/Apr/16&+0.5& 212.30& -43.98& 317.42& 16.68\\
2007/Apr/17&-0.5& 206.97& -44.98& 313.31& 16.74\\
2007/Apr/17&+0.5& 206.97& -43.98& 313.55& 17.72\\
2007/Apr/17&+0.5& 204.97& -43.98& 312.07& 18.02\\
2007/Apr/17&-0.5& 204.97& -44.98& 311.87& 17.04\\
2007/Apr/19&-0.5& 209.80& -44.98& 315.33& 16.25\\
2007/Apr/19&+0.5& 209.80& -43.98& 315.62& 17.21\\
2007/Apr/21&+0.5& 197.26& -43.98& 306.29& 18.77\\
2007/Apr/21&-0.5& 197.26& -44.98& 306.21& 17.77\\
2007/Apr/22&-0.5& 202.25& -44.98& 309.89& 17.38\\
2007/Apr/22&+0.5& 202.25& -43.98& 310.05& 18.36\\
2007/May/12&-0.5& 237.43& -44.98& 332.96& 7.24\\
2007/Mar/12&+0.5& 237.43& -43.98& 333.60& 8.02\\
2007/May/15&+0.5& 197.43& -43.98& 306.41& 18.76\\
2007/May/15&-0.5& 197.43& -44.98& 306.33& 17.76\\
2007/May/16&-0.5& 189.93& -44.98& 300.76& 17.84\\
2007/May/16&+0.5& 189.93& -43.98& 300.71& 18.83\\
2007/May/17&+0.5& 221.18& -43.98& 323.56& 14.30\\
2007/May/17&-0.5& 221.18& -44.98& 323.12& 13.40\\
2007/May/20&+0.5& 214.93& -43.98& 319.28& 16.06\\
2007/May/20&-0.5& 214.93& -44.98& 318.92& 15.13\\
2007/May/20&+0.5& 279.93& -43.98& 351.48& -16.50\\
2007/May/20&-0.5& 279.93& -44.98& 350.51& -16.87\\
2007/May/21&+0.5& 227.43& -43.98& 327.62& 12.16\\
2007/May/21&-0.5& 227.43& -44.98& 327.10& 11.30\\
2007/May/21&+0.5& 282.43& -43.98& 352.16& -18.18\\
2007/May/21&-0.5& 282.43& -44.98& 351.18& -18.53\\
2007/Jun/13&+0.5& 219.93& -43.98& 322.72& 14.68\\
2007/Jun/13&-0.5& 219.93& -44.98& 322.29& 13.78\\
2007/Jun/13&-0.5& 288.63& -44.98& 352.64& -22.69\\
2007/Jun/13&+0.5& 288.63& -43.98& 353.68& -22.41\\
2007/Jun/14&-0.5& 220.08& -43.98& 322.83& 14.64\\
2007/Jun/14&-0.5& 220.08& -44.98& 322.40& 13.73\\
2007/Jun/14&-0.5& 288.86& -44.98& 352.69& -22.85\\
2007/Jun/14&+0.5& 288.86& -43.98& 353.73& -22.57\\
2007/Jun/15&+0.5& 220.08& -43.98& 322.83& 14.64\\
2007/Jun/15&-0.5& 220.08& -44.98& 322.40& 13.73\\
2007/Jun/15&-0.5& 288.86& -44.98& 352.69& -22.85\\
2007/Jun/15&+0.5& 288.86& -43.98& 353.73& -22.57\\
\hline
\end{tabular}
\end{table}
Each night was divided into two or three series of observation modes,  
such as $\lq\lq$ON--OFF",
$\lq\lq$OFF--ON--OFF", or $\lq\lq$OFF--ON" observations. 
The ON-source run was scheduled  
to contain the meridian passage of the target, and 
 the OFF-source run to follow the zenith angle distribution 
 of the ON-source observation.
On the average, 
the OFF source region had an offset in the R.A. of
more than 30$^\circ$ away from
the target. 

In the 2004 observation,
the data of each telescope of T2, T3 and T4 were recorded with GPS time stamps, 
independently 
from the other telescope, when
more than four photomultiplier (PMT) signals 
exceeded 7.6 p.e.(photo-electrons) in each telescope. 
In the stage of offline analysis, 
the GPS time was used as a token to find coincidence events 
for the three telescopes. 
In the 2007 observation, a trigger circuit was employed 
into the electronics system \citep{nishijima} 
so that the events of more than two telescopes in coincidence  
were selected to be recorded. 

The total efficiency of light collection, including the reflectivity
of the segmented mirrors, the light guides, and the quantum efficiency 
of the photomultiplier tubes, was monitored 
by ``muon-ring analysis"  of the Cherenkov lights radiated 
by cosmic-ray muons \citep{enomoto_vela},  
during the period of the observation. 
The deterioration ratios of the reflectivity relative to the initial value 
when the mirrors were made were
measured to be 0.55, 0.60, and 0.75, for T2, T3, and T4, 
respectively, in 2004. 
The telescope T4 was built recently,  
and show the highest value of reflectivity.
The deterioration rate was found to be slightly higher than the
 10\% level per
year. 
Before the 2007 observation was commenced, 
 mirror cleaning work using water was
carried out.  
The reflectivity was, however, not recovered 
to the initial value. 
The reflectivity of T2 became too low to be known as being available 
from the calibration data of the muon ring.   
The two-fold coincidence data (i.e., T3 and T4) were 
used in the 2007 analysis.

We selected those events in which 
the images of the Cherenkov light consisted 
of at least five adjacent pixels exceeding the 5\,p.e.\ threshold 
(called the $\lq\lq$cluster events"). 
For the 2004 runs,
the frequency of the three-fold coincidence was 8$\sim$10 Hz,  
giving the rate of the cluster events as 6$\sim$7\,Hz.  
In 2007, the two-fold coincidence between  T3 and T4 gave  
a trigger rate of 12$\sim$14 Hz, which was   
reduced to 6$\sim$7 Hz for the cluster events.
A cloud in the sky caused a
low trigger rate when it blocked the direction of the telescope pointed, 
and events during the time of such a low-rate trigger 
were excluded from the analysis. 
The effective observation times 
for the 2004 and 2007 observations were 
996 (998) and 2187 (2386) minutes, respectively, 
where the time for an OFF-source run is shown in parentheses. 
The mean of the zenith angle of the data used for the analysis 
was 19.8$^\circ$ 
and 18.6$^\circ$, respectively.
 
\section{Method of Analysis}

Since the ON-source run sees    
 the Galactic plane and the OFF-source direction 
is away from the plane, as described in Table \ref{posoff}, 
the amount of night sky background (NSB) light 
differs between the ON- and OFF-source data.
The region of the sky for the ON-source run is brighter than 
the OFF case. 
This may cause a spurious effect of having excess events 
created by star light along the Milky Way,   
when we subtract the number of OFF-source events 
from ON-source ones.

The NSB affects  
the distribution of ADC (Analog to Digital Converter) counts 
in the electronics circuit, and in turn,  
the degree of NSB is monitored in each observation run. 
Darker NSB in an OFF-source observation 
can be $\lq\lq$corrected for" to match the NSB in the case of ON-source run,  
by applying the method of $\lq\lq$software padding"  \citep{padding}   
to add extra background light onto the photomultiplier tubes of 
OFF-source data. 
%%%%%%%%%%%%%%%'±'±'©'ç%%%%%%%%%%%%%%%% 
We applied a somewhat simplified software padding, 
when compared with the method developed by the Whipple group\citep{padding}. 
Extra background lights of a common shape of Gaussian fluctuation are added 
to ADC data of all PMTs for all OFF-source data in common. 
This padding procedure
was applied only to the PMTs of having a signal of TDC (Time-to-Digital Converter) hit,  
since the CANGAROO-III analysis method requires TDC timing
information.
%%%%%%%%%%%%%%%'±'±'Ü'Å%%%%%%%%%%%%%%%%%%  

The effect that may be caused by adding more NSB as an artifact   
to the pixels of the photomultiplier tubes was studied 
by a Monte-Carlo simulation. 
It was not easy to precisely estimate and correct for the effects.   
However, 
the spurious excess events could be made not to be effective  
by raising the threshold for  
discriminating the Cherenkov light signal in each pixel of the camera.  
We applied a threshold of as high as 8 or 10 p.e. for 2004 or 2006 data,
respectively,
to each photomultiplier pixel    
 of the clustering Cherenkov light image. 

The analysis procedures used in this work were basically the same as those 
described in \citet{cena}, but with more details to be found 
in \citet{enomoto_vela} and \citet{enomoto_0852}.
As the first step, the Hillas parameters of a 
Cherenkov light image \citep{hillas} 
were calculated for three or two telescopes.
The arrival direction   
was determined by the condition of  minimizing 
the sum of the squared widths ($\chi^2_0$; weighted by the photon yield) 
of the images seen from the assumed arrival position, 
in addition by putting a constraint on the $Distance$ from the arrival position  
to the center of image.

As a measure of gamma-ray likeliness of each event,
we used  
the Fisher Discriminant (hereafter called as $FD$) \citep{fisher,enomoto_vela}, 
which is given by a linear combination of 
the input parameters, as    
$$FD = \sum \alpha_i \cdot P_i.$$
The coefficients $\alpha_i$ were chosen   
to realize the best separation between the gamma-ray and background events, 
where $P_i$ is a component of vector, $\vec{P}=(W2,W3,W4,L2,L3,L4)$ 
or $\vec{P}=(W3,W4,L3,L4)$,  
 each of $W2,W3,W4,L2,L3,L4$ being the energy-corrected 
$Widths$ or $Lengths$ 
observed with telescopes 
T2, T3, and T4; for details see a reference \citep{enomoto_vela}.
We excluded from analysis  
those events in which the photomultiplier pixels 
in the outermost layer of the camera 
were hit by photons ($\lq\lq$edge cut"), 
because the image of Cherenkov light of the events was deformed,  
particularly in the distribution of the $Length$  parameter.  

We estimated 
the response function, $f_{\gamma}$, of $FD$ for gamma-rays from  
a Monte-Carlo simulation, and obtained   
the function $f_{b}$ for background events from     
the distribution of the OFF-source data.  
Note that the energy spectrum 
proportional to $E^\gamma$ was used for gamma-rays 
in the Monte-Carlo simulation,   
where $E$ is gamma-ray energy and $\gamma$=$-$2.1. 
Then,  
the observed distribution, $F_{on}$,
 as a function of $FD$ for ON-source data  
 was equated to   
a linear combination of $f_{\gamma}$ and $f_{b}$, as   
$$  F_{on}=N_b f_{b} + N_{\gamma} f_{\gamma}= F_b + F_{\gamma}.$$ 
The number of background and gamma-ray events contained 
in the ON-source data were designated as $N_b$ and $N_{\gamma}$, and  
were determined by fitting the $FD$ distribution of the ON-source data 
to $N_b f_{b} + N_{\gamma} f_{\gamma}$
under the condition of $N_{on}= N_b + N_{\gamma}$,   
where $N_{on}$ is the total number of events of the ON-source data.  
The functions $F_b=N_bf_b$ and $F_{\gamma}=N_{\gamma}f_{\gamma}$ represent 
the distribution of the 
background and gamma-ray events with the total number 
of $N_b$ and $N_{\gamma}$, respectively.  

The response function ($f_{\gamma}$ and $f_b$) depends on 
its location in the field of view of the telescope, and thus   
a fitting for ON-source data 
by the sum of contributions from $f_{\gamma}$ and $f_{b}$ 
was carried out for every arrival direction investigated, 
 within the field of view of the telescope.  
This way of inferring $N_{\gamma}$, the number of gamma-ray events, 
was adopted when we used the method (a).     

The method (a) of inferring $N_{\gamma}$ 
by fitting the $FD$ distribution was 
applied to the data of CANGAROO-III,  
to bring about the detection of gamma-ray signals successfully   
from point-like sources \citep{sakamoto,nakamori} as well as  
from a source of considerable extension \citep{enomoto_0852}, 
and also to set upper limits on gamma-rays from other objects \citep{cena,enomoto_1987a}.
Note that with this method of fitting $FD$, 
it was not required to normalize the number of events 
by the observation time, 
in the case that the observation times of ON- and OFF-source runs 
were different  from each other. 

A Monte Carlo simulation shows that 
gamma-ray events are likely to have $FD$ larger than $-0.5$, 
which is consistent with the gamma-ray candidate events 
extracted from the observation data by using method (a),   
as is presented in the next section. 
The events of $FD$ larger than $-0.5$ were 
selected as gamma-ray enriched ones   
in the case of method (b):   
the wobble and ring background method.    

\section{Results with the method of ON and OFF runs: Method (a)}

\begin{figure}[htbp]
  \plotone{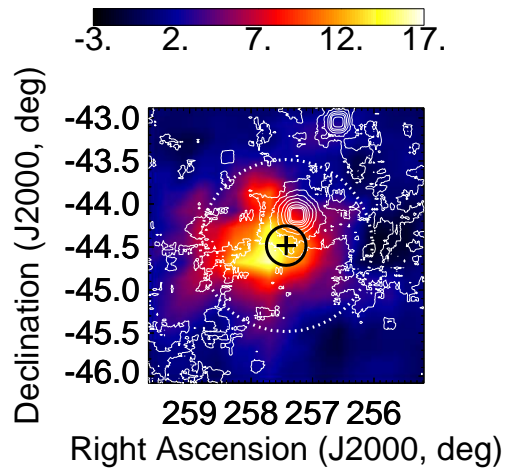}
  \caption{Map of excess events of 2004 data by using method (a) 
of the ON- and OFF source observations. 
The number of excess events per 0.2$^o$ $\times$ 0.2$^o$ cell  
is plotted in the equatorial coordinate. 
 The number of excess events was smoothed 
by taking the average of the 3 $\times$ 3 = 9 cells 
around each investigated direction.  
The black cross at the center of the map indicated 
the position of the pulsar PSR B1706-44, 
the radius of the black circle represents the point spread function (PSF) 
of $\delta\theta_0=0.24^o$,
and the circle of white dotted line shows the region within 
$1^o$ radius from the pulsar. 
The inserted white contours 
are the hard-X-ray map obtained from the ROSAT satellite \citep{skyview}.}
  \label{emap2004}
\end{figure}

The arrival directions in the equatorial coordinate (J2000) 
of observed events were sorted   
 into 19 $\times$ 19=361 cells with
each having a 0.2$^o$ $\times$ 0.2$^o$ size   
in a field of view of   
$\Omega_{FoV}= 3.8^o\times 3.8^o$.  
For each cell of the arrival direction, 
the $FD$ distribution of ON- and OFF-source runs 
was constructed to yield  
the parameter $N_{\gamma}$, {\it i.e.} 
the number of gamma-ray events,   
giving a map of 
 gamma-ray-like events of the 2004 data, as plotted in Fig. \ref{emap2004}.
Smoothing of taking the average of the neighboring 3 $\times$ 3 cells 
  was carried out for each investigated direction 
in the morphology map of the figure. 
The black cross at the center of the map indicates the position 
of the pulsar PSR B1706$-$44, 
the radius of the black circle shows the point spread function (PSF) of 
$\delta\theta_0=0.24^o$,
and the circle drawn with  
the white dotted line shows the region within   
$1^o$ distance from the pulsar.  % ({\bf fiducial volume ??}). 
The white contours  
are the ROSAT hard X-ray map taken from NASA Skyview \citep{skyview}.  
 
The excess, 
which is much larger than 
the statistical fluctuation of    
$\sim \pm 2$ events per 0.2$^o$ $\times$ 0.2$^o$ area,  
shows a broader distribution than the point spread function. 
No prominent point-like excess appears 
either at the position of the pulsar PSR~B1706-44  
or at any other directions in the vicinity of the pulsar,     
which is consistent with the H.E.S.S. observation \citep{hess} 
of no detection of point-like emission 
from the pulsar PSR~B1706$-$44. 

\begin{figure}[htbp]
  \plotone{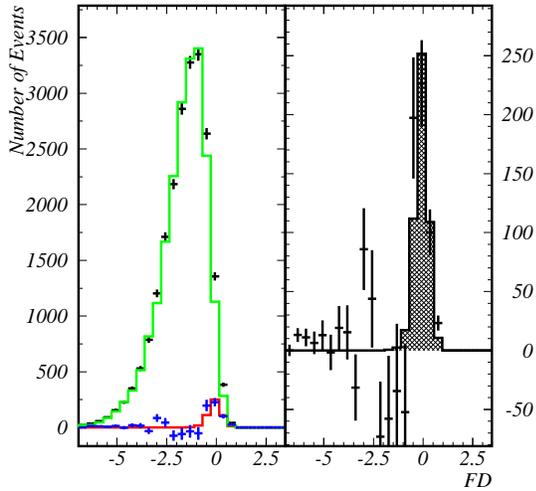}
  \caption{$FD$ distributions within 1.0$^o$ from PSR B1706-44,  
obtained from 2004 data by using method (a) 
of ON- and OFF- source observations. 
In the left panel, the data points with error bars (statistical errors)
are the ON-source run, {\it i.e.,}$F_{on}$ by the black cross.  
The green histogram represents $F_b$, the
 OFF-source data,
the crosses of blue color are the excess counts of $F_{on} - F_b$,  
and the histogram of red color shows $F_{\gamma}$,  
the response of gamma-rays ($f_{\gamma}$) inferred from 
a Monte-Carlo simulation 
multiplied by $N_{\gamma}$. 
The right panel shows a magnified view of the excess events.} 
  \label{fd2004}
\end{figure}

 The $FD$ distribution of    
 the observed events  within a radius of  1.0$^o$ 
from PSR B1706$-$44 is shown in Fig. \ref{fd2004}.  
The number of events of the ON-source data 
(the black cross) exceeds the distribution 
of the OFF-source events $F_b$ (the green histogram) around $FD~\sim~0$. 
The excess events, $(F_{on}-F_b)$,  
 indicated by the blue cross    
concentrate in $FD =-0.5$ to $0.5$,    
is consistent with the gamma-ray distribution, $F_{\gamma}$, inferred from  
the Monte-Carlo simulation and shown by the blue histogram. 
The right panel shows a magnified view of $F_{\gamma}$ and 
$(F_{on} - F_b)$. 
The number of excess deduced is 504 $\pm$ 82
(6.1 $\sigma$). 

\begin{figure}[htbp]
  \epsscale{.80} 
  \plotone{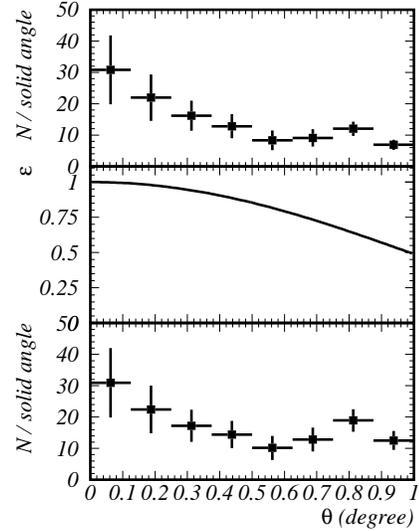}
  \caption{Excess count map per solid angle (arbitrary unit)
in $\theta$ space (the upper panel).
Result of 2004 data by using method (a) 
of ON- and OFF-source observations. 
The middle panel is the estimated acceptance normalized at $\theta$
at zero and the bottom is the acceptance corrected $\theta$
distribution. The vertical axis is in arbitrary unit.}
  \label{q2004}
\end{figure}

\begin{figure}[htbp]
  \plotone{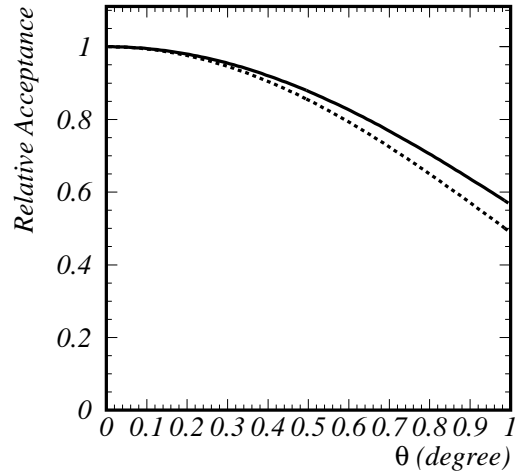}
  \caption{Relative acceptance versus angular distance from the center of the field
of view. 
The dashed line is for the three-fold analysis
and the solid for the two-fold one. 
% the both curves normalized as 1 at $\theta$ = 0.
}
  \label{q2acc}
\end{figure}

The map of Fig. \ref{emap2004} might give an impression  
that the angular extent of the extended emission observed 
can be definitely determined to be as large as $1^o$. 
However, we have a difficulty to make a conclusive argument 
about the angular extent beyond $1^o$, 
since  the $\lq\lq$effective" field of view of our observation 
was limited to within about 1$^o$ (half of the camera radius), 
due to the effect that a Cherenkov light image is deformed 
when a part of it is 
located near the outer edge of the camera. 
In order to estimate the angular extent of the excess events, 
we plotted in the top panel of Fig. \ref{q2004} 
the distribution of gamma-ray-like events versus $\theta$, 
where $\theta$ is the angular distance of the arrival direction 
of each event from the black cross  
at the center of the map,  the position of PSR B1706$-$44. 
The detection efficiency as a function of the angle 
from the center of the field of view was estimated 
by a Monte-Carlo simulation, and also by using the number 
of gamma-ray-like events of $FD >-0.5$ in an  OFF-source run. 
As presented by the plot in the middle panel of Fig. \ref{q2004}, 
the efficiency decreases with $\theta$, 
due mainly to the $\lq\lq$edge cut".  
The acceptance of detection differs slightly between the 
2004 and 2007 year observations, as shown by Fig. \ref{q2acc}. 
The acceptance for the 2007 data 
(the solid line) is wider than 
the 2004 one (the dashed line),  
due to the two-fold analysis for 2007 data 
when compared with the case of the three-fold analysis for 2004 data.

\begin{figure}[htbp]
  \plotone{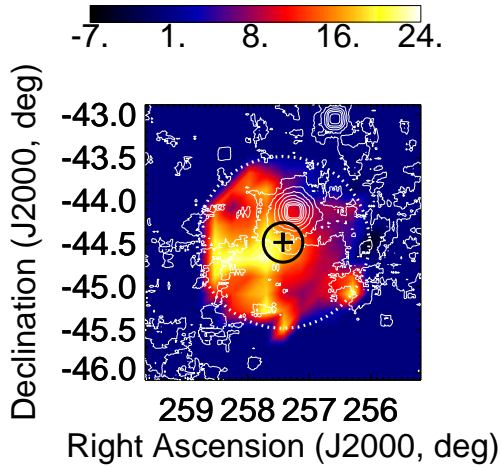}
  \caption{Map of excess events of 2004 data by using method (a) 
of ON- and OFF-source observations after  
an acceptance correction. 
The definitions are the same with those in Fig. \ref{emap2004}.}
  \label{mor2004}
\end{figure}

The $\theta$ distribution, 
after being corrected for the detection efficiency, 
is shown in the bottom panel of Fig. \ref{q2004}. 
A correction for the acceptance   
was also applied to the morphology map of 
 Fig. \ref{emap2004}, 
and is presented in Fig. \ref{mor2004}.
We limited the plot of data in this figure 
%Fig. \ref{mor2004}  of the acceptance corrected   
to be within a region, less than 1$^o$ from PSR B1706$-$44. 
The acceptance decreases with $\theta$  rapidly outside of $1^o$, 
%as shown in the middle panel of Fig. \ref{q2004},   
and  the data seriously suffer  
from statistical fluctuation and systematical error 
at $\theta > 1^o$  

\begin{figure*}[htbp]
\includegraphics[width=200pt]{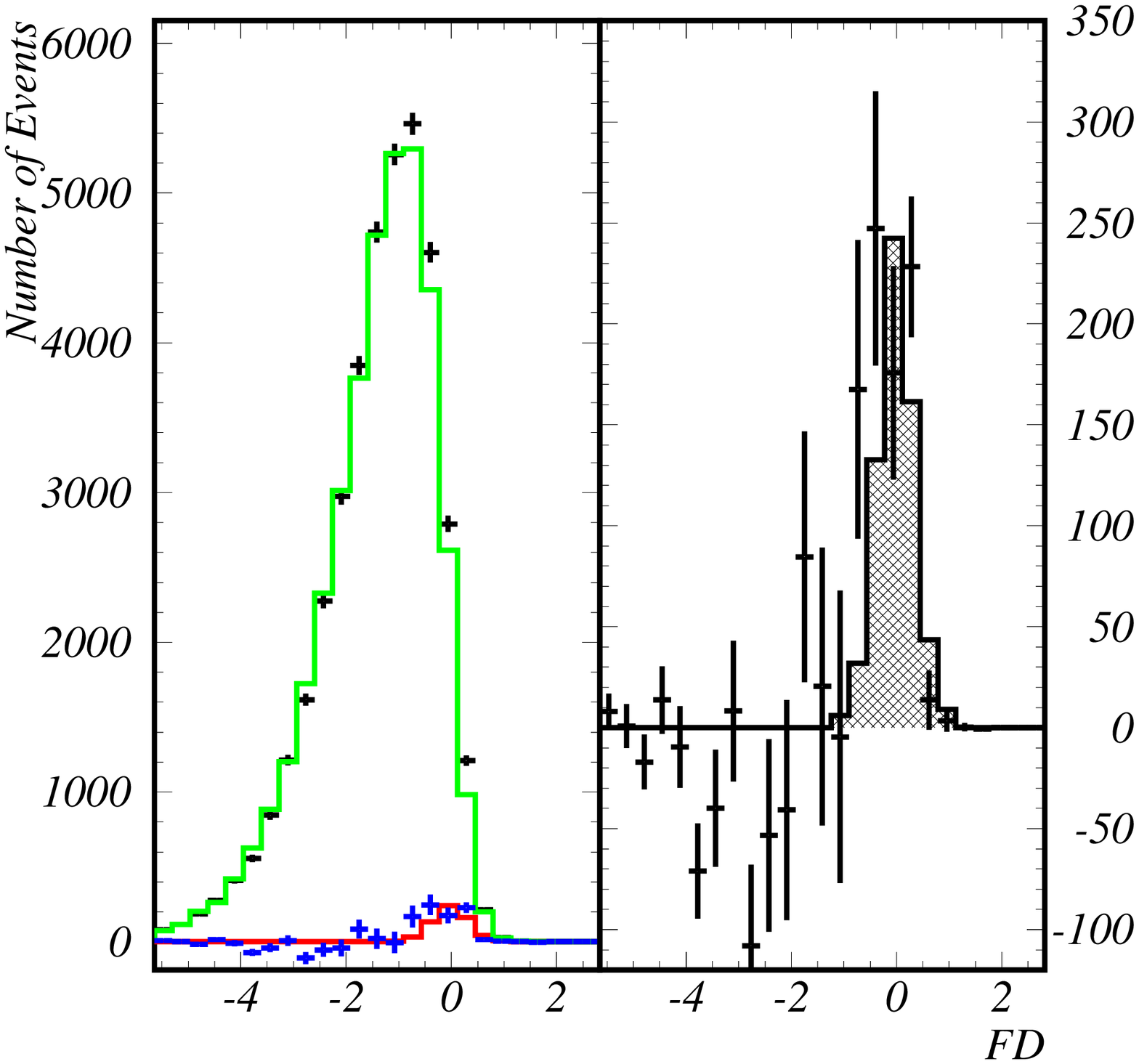}
\hskip 2.0cm
\includegraphics[width=200pt]{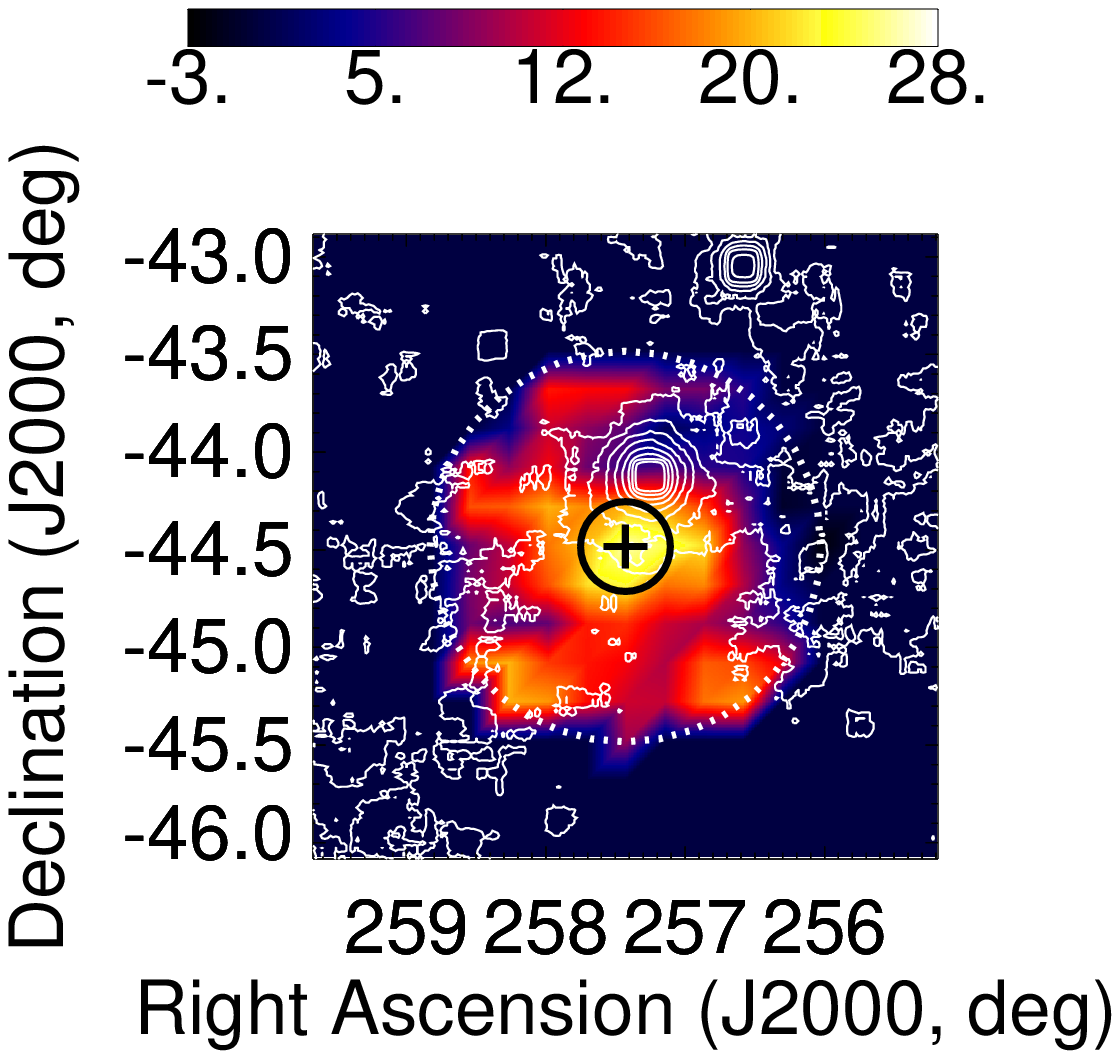}
\caption{$FD$ distribution (left panel) and acceptance corrected
excess count map (right panel) for the 2007 data  
by using method (a) of ON- and OFF-source observations.
The notations are the same as in Fig. \ref{fd2004} and
\ref{mor2004}.}
\label{results2007}
\end{figure*}

The result from data taken in
2007 is shown in Fig. \ref{results2007},   
 the $FD$ distribution is in the left panel 
and the morphology map of the arrival direction is in the right,  
the number of gamma-ray-like events 
of the ON-source over the OFF-source runs 
is 627 $\pm$ 127 events for the region within $\theta<1.0^o$
(5.1 $\sigma$). 

\begin{figure}[htbp]
  \plotone{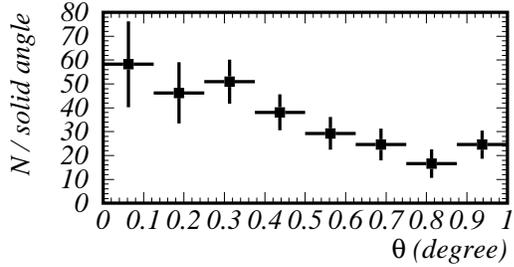}
  \caption{Acceptance-corrected $\theta$ plot from the excess events of 
2004 and 2007 data combined.
  }
  \label{qcomb}
\end{figure}

\begin{figure}[htbp]
  \plotone{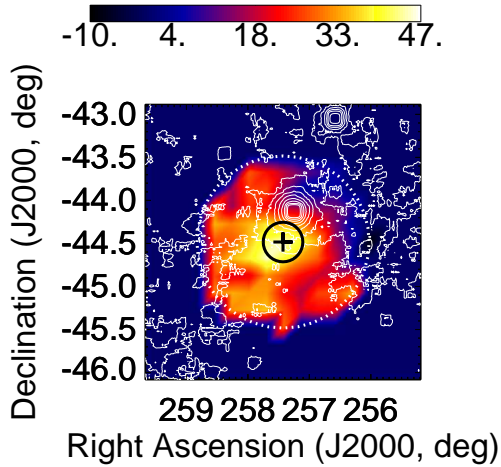}
  \caption{Acceptance-corrected morphology map of excess events from  
2004 and 2007 data combined.
The definitions of contours and circle are the same as in Fig. \ref{mor2004}.
}
  \label{moradd}
\end{figure}

The $\theta$ plot from the combined data of years 2004 and 2007
after the acceptance correction is shown in Fig. \ref{qcomb}.
The $\chi^2$/d.o.f value (the chi-square value per degree of freedom) 
calculated for flat distribution, 
{\it i.e.} the case of constant excess counts against $\theta$, is 16.8/7. 
When we fit the distribution  
by a test function $\propto$ 1-$\theta$/a from the pulsar position, 
the best fit for the observed distribution 
was obtained when $a=1.36^o \pm 0.21^o$ with $\chi^2$/d.o.f. = 3.7/6.
In the case of fitting by a Gaussian function $\propto exp(-\theta^2/(2\sigma^2))$, 
the $\chi^2$/d.o.f value was 4.0/6, with
$\sigma =0.62^o \pm 0.10^o$. 
Fitting was also made for
 the case of $\lq\lq$constant value plus Gaussian function",  
{\it i.e.} $\propto N_c + a \cdot exp(-\theta^2/(2\sigma^2))$, 
yielding results of $N_c = 19\pm6$, $a=40\pm11$, 
and $\sigma =0.34^o \pm 0.10^o$ with $\chi^2$/d.o.f equal to $1.9/5$. 
A morphology map for the combined data of years 2004 and 2007 
after the acceptance correction is shown in Fig. \ref{moradd}.

\begin{figure}[htbp]
  \plotone{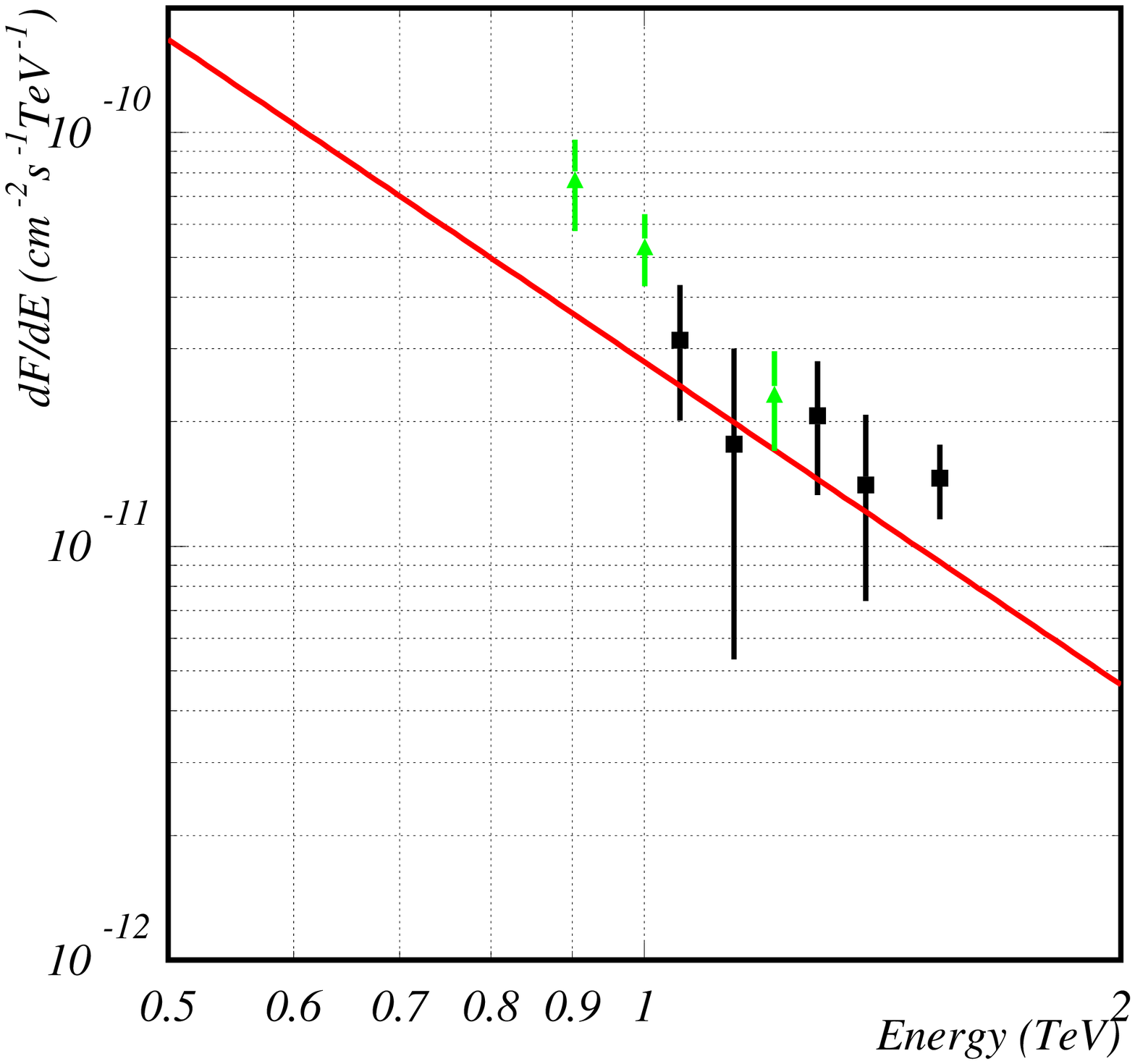}
  \caption{Differential flux from the area within $1.0^o$ radius 
 (solid angle of $9.6 \times 10^{-4}$) around PSR~B1706$-$44. 
The green points are from the
2004 runs and the black from the 2007 runs.
The errors are statistical ones. 
The red line indicates the flux from Crab Nebula 
observed by HEGRA \citep{hegra}.  
}
  \label{ftev}
\end{figure}

The flux integrated within $1.0^o$    
from PSR~B1706$-$44 is plotted 
in Fig. \ref{ftev} against gamma-ray energy $E$.
The green points indicate the 2004 data and the
black ones the 2007 data, respectively,  
with the error bars showing the statistical fluctuation.  
The flux of the Crab Nebula, reported by HEGRA \citep{hegra}, 
is shown by the red line for a comparison. 
The flux is given numerically   
as a function of $E$ in Table \ref{tab_fwide}.
\begin{table}
\caption{Differential Flux from the area of $1.0^o$ radius 
 (corresponding to $9.57\times 10^{-4}$ sr) around PSR~B1706$-$44.}
\label{tab_fwide}
\begin{tabular}{cccc}
\hline\hline
Epoch & $<$E$>$ & dF/dE & $\Delta$(dF/dE) \\
year & TeV & 10$^{-11}$cm$^{-2}$s$^{-1}$TeV$^{-1}$ & 
10$^{-11}$cm$^{-2}$s$^{-1}$TeV$^{-1}$ \\
\hline
2004 
& 0.904  &   7.7 & 1.9\\
&  1.00  &   5.3 & 1.0\\
&  1.21  &   2.3 & 0.6\\
2007 
&  1.05  &   3.1 & 1.1\\
&  1.14  &   1.8 & 1.2\\
&  1.29  &   2.1 & 0.7\\
&  1.38  &   1.4 & 0.7\\
&  1.54  &   1.5 & 0.3\\
\hline\hline
\end{tabular}
\end{table}

In addition to our standard systematic errors such as on
energy determination (photon collection efficiencies), we investigated
the possibility of systematic difference of $FD$ (gamma-ray likeliness)
between two data obtained in different pointings.
The $width$ and $length$ (image parameter) have a dependence on
the elevation angle. The image becomes smaller in smaller elevation.
Without any correction, the response of $FD$ for background events
 in the ON observation 
sometimes differ from that of the OFF. This is mainly
due to the difference in
the ON/OFF distributions of the elevation angle. After a correction, however,
this systematic error is reduced which was proved by the analysis
of the off-Galactic-Plane data. 
The systematic error level is considered to be
less than 5\%.
The bright optical lights along the Galactic plane can deform 
 the image parameters.   
We checked the $FD$ responses by increasing NSB in the Monte-Carlo
simulations and the OFF data (by software padding). 
When NSB increases, small sized images which look like
gamma-ray images increase. 
These images can be rejected by increasing the threshold
of pixel. This is a reason why we used 10 p.e. threshold while the standard
analysis used 5 p.e. We increased NSB amount and/or decrease this pixel
threshold and check the differences in the obtained gamma-ray fluxes.
We, however, found that the main systematic error are due to NSB. 
The flat
component observed in this analysis still remain to be statistically
significant even in the worst case. The fluctuations in the
observed flux changes approximately 30\% level. 
Our claim for the
flat component is, therefore, marginal. 
The resulting uncertainties in gamma-ray fluxes is  +47, and -37\%. 

\section{Results with Wobble and the Ring Background Method: Method (b)}
 
\begin{figure}[htbp]
  \plotone{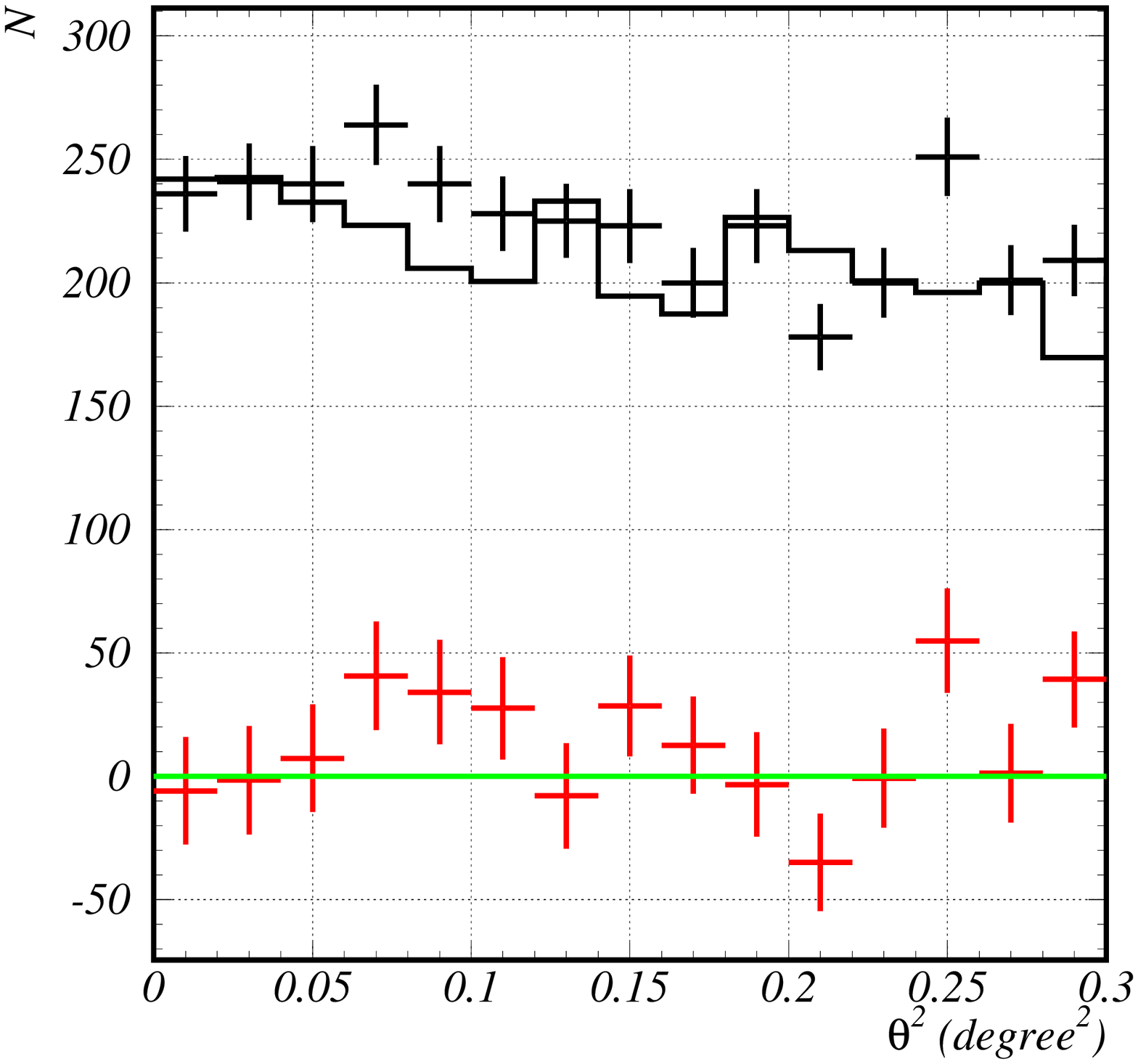}
  \caption{$\theta^2$ distributions of the 2007 data 
obtained by method (b1), the wobble method. 
The black crosses indicate the number of events  
 obtained as the $\lq\lq$ON-source counts". 
The black histogram shows that obtained as the background number of events.
The red points with error bars are the subtracted data of 
$\lq\lq$ON-source" minus the emulated background events, 
{\it i.e.} the $\lq\lq$off-source" events. 
The green line indicates the zero position of the number of events. 
}
  \label{w2007}
\end{figure}

To the 2007 data, which were taken in the wobble mode observation, 
the $\lq\lq$wobble" and 
the $\lq\lq$ring background method" can be directly applied    
in a way similar to what is  used by the H.E.S.S. Group.    
The arrival directions along the circumference of its radius equal to 
$\theta_{r1}=0.5^o$ 
from the center of the field of view are guaranteed 
to have the same, uniform detection efficiency.   
Along the circle of $\theta_{r1}=0.5^o$, 
six points that are successively at every $60^o$ of the opening angle 
around the center of the field of view 
were selected as the central points of $\theta=0$ for calculating the $\theta^2$ 
distribution of Fig. \ref{w2007}, 
where $\theta$ is the angular distance from PSR~B1706$-$44. 
One of the six points corresponds to the center of $\lq\lq$ON-source" directions, 
 {\it i.e.} at the position of PSR~B1706$-$44. 
The other five points were utilized as the $\lq\lq$off-source direction"   
 for emulating background events.   
The events of having $FD~>~-0.5$ were 
selected as gamma-rays enriched events, 
and then  
the numbers of events from $\lq\lq$ON-" and $\lq\lq$off-source" directions were compared 
 for estimating the flux of gamma-rays 
as a function of $\theta^2$.  
We call  
this method of using the circle of radius $\theta_{r1}=0.5^o$ 
as method (b1).  

Fig. \ref{w2007} shows  
the $\theta^2$ distributions from the 2007 data
by taking the events of $FD > -0.5$. 
When the cut  
$\theta^2 <0.06$ degree$^2$ was applied to the observed $\theta^2$  
distribution, 
as matched to the point-spread function,   
the excess was below 100 events,  
giving the upper bound of the flux roughly lower than the level of 20\% Crab.

\begin{figure}[htbp]
  \plotone{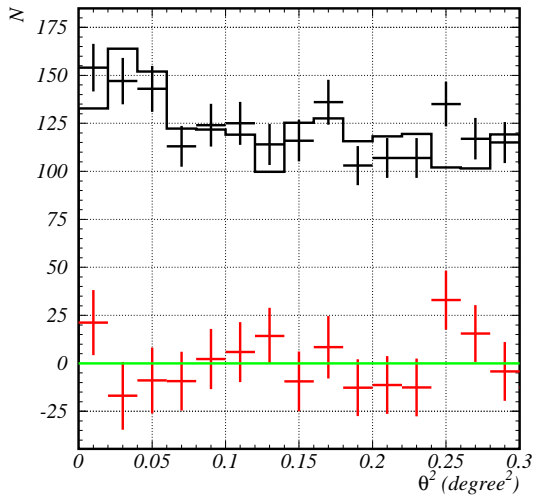}
  \caption{$\theta^2$ distributions of the 2004 data 
obtained by method (b1); the wobble method. 
The notations are the same as those in Fig. \ref{w2007}.  
}
  \label{w2004}
\end{figure}

To the 2004 data, method (b1) could not be directly applied, 
because the points along the circumference of $\theta_{r1}=0.5^o$ radius,    
%$\lq\lq$a wobble-ring region" along the   
on which the direction of PSR~B1706$-$44 is located, 
 did not have a constant, uniform efficiency of acceptance.
It was necessary to correct for the varying acceptance  
by using a Monte-Carlo simulation.  
The $\theta^2$ distribution, 
after the number of events was corrected for 
 is shown in Fig. \ref{w2004}.
There appears to be no statistically significant evidence 
of point-like gamma-ray emission.  
The upper limit was set at the position of the pulsar 
PSR~B1706$-$44 to be less than 20\% Crab flux.

\begin{figure}[htbp]
  \plotone{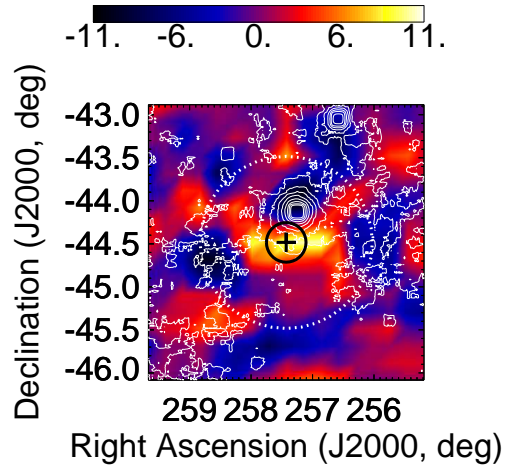}
  \caption{Morphology map of excess events 
for the 2007 data 
obtained by method (b2) of the ring background method. 
Details of the analysis are to be found in the text.}
  \label{hess2007}
\end{figure}

\begin{figure}[htbp]
  \plotone{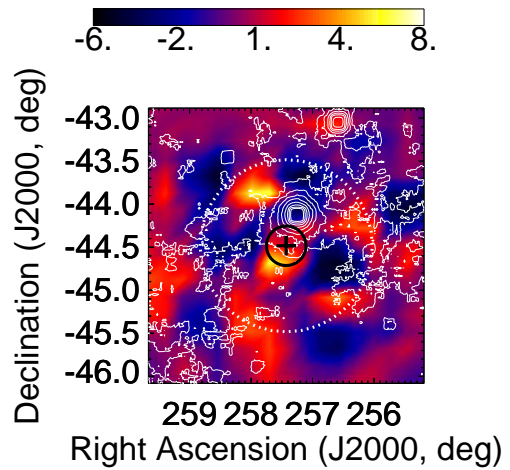}
  \caption{Morphology map of excess events obtained from the 2004 data 
obtained by method (b2) of the ring background method. 
Details of the analysis are to be found in the text.
}
  \label{hess2004}
\end{figure}

For constructing a morphology map, 
we used 19 $\times$ 19 $FD$ histograms of ON-source data, 
each of which was constructed from the events falling in 
a 0.2$^o$ $\times$ 0.2$^o$ area,   
within the field of view of $\Omega_{FoV}=3.8^o\times3.8^o$.   
The background was estimated from those events, 
of which the arrival direction is at distance $\theta_{r2}$ 
from the investigated direction, namely within 
 a ring-shaped region between $\theta_{r2} =0.40^o$ and $0.55^o$ 
from the investigated direction.  
We call this way of estimating background events 
method (b2) of the ring background. 
It is noted that this method differs slightly 
from $\lq\lq$the H.E.S.S. method",   
in using different values for some parameters such as 
the angular width. 
Background subtraction was then carried out 
after an acceptance correction. 
A map of excess events is plotted in Fig. \ref{hess2007}.
%The color bar at the top of Fig. \ref{hess2007} locates 
%the number of entries in the cell of 
% 0.2$^o$ $\times$ 0.2$^o$ for the counts between $-$11 and 11 
% which correspond to the minimum and the maximum number of 
%the entries to the cells. 
The distribution of excess events is consistent with the statistical
fluctuation, with no indication of point-like source seen.
The result of applying method (b2) to the 2004 data  
is presented in Fig. \ref{hess2004}. 
No prominent excess appears beyond the statistical fluctuation 
in the morphology map around PSR~B1706-44. 

\section{Discussions}

The method (a) of  
having two separate observations of ON- and OFF-source runs  
has indicated emission of TeV gamma-rays 
%provided us with an indication of    
from an extended region around the pulsar PSR~B1706$-$44.   
On the other hand, 
method (b) of either the wobble or the ring background method,  
which is based on the subtraction of background events 
by using the neighboring region 
displaced by an angular distance of $\sim 0.5^o$,     
could not confirm the result by method (a). 
The excess observed from method (b) is consistent 
with the statistical fluctuation, 
although arguments could be made about some spatial correlation 
that may exist in the morphological map of excess counts. 

The two differing results obtained 
through the two methods,  
both of which were applied to the same ON-source data, 
can, however, be understood as not in contradiction with each other,      
if we admit that gamma-ray events are contained in the dataset 
which is utilized for emulating background events.  
%the technique of IACT  still has a
% limitation in distinguishing gamma rays from the cosmic-ray background.  
 
As can be seen in Fig. \ref{fd2004}, the candidate events for gamma-rays  
are only $\sim$ 2\% of 
the total number of ON-source events. 
Such a small ratio of the signal-to-noise demands us    
 to subtract from the ON-source data such events      
that are emulated from observed data containing presumably no gamma-ray events  
 and are 
regarded as being equivalent to background events in the ON-source data.   
However, if a considerable amount of gamma-ray 
events are included in the dataset for emulating background,  
the subtraction    
can diminish and even kill a gamma-ray signal that may exist.  
Such a chance may be more likely to occur in method (b)  
than in the case of method (a) of using 
the data from the region off the Galactic disk.    
  
%However, 
Diffuse gamma-rays from the Galactic disk should be subtracted 
to determine the genuine intensity of individual sources, and thus 
the method (b) might be preferable to measure the flux    
from a source embedded in the Galactic disk. 
%implying that method (b) seems to be preferable  
%for knowing the genuine flux from a gamma-ray source.  
In fact, 
the diffuse gamma rays from the Galactic disk 
is expected to have a flux as large as   
$F_{disk}=10^{-8}\sim 10^{-7}$ cm$^{-2}$s$^{-1}$TeV$^{-1}$sr$^{-1}$ at 1TeV, 
when extrapolated from the 10 GeV energy region, or taking the case of  
 the Galactic Center region \citep{hess_gp}. 
Our result for the flux within an area of $1^o$ radius 
is $2.2 \times 10^{-11}$ cm$^{-2}$s$^{-1}$, 
as large as the intensity of the Crab nebula.  
The flux $(F_{disk})$, when integrated over solid angle  
 of a $0.2^o\times0.2^o = 3.8\times 10^{-5}$sr,   
amounts to 
$10^{-13}\sim 10^{-12}$cm$^{-2}$s$^{-1}$TeV$^{-1}$, 
which is equivalent to 0.01 $\sim$ 0.1 of the flux from the Crab nebula.  
This intensity is similar to the flux from Galactic sources that   
 the scan survey of the Galactic plane uncovered \citep{hess_science}.

If the emission from the Galactic disk is adequately uniform,
 or in other words, 
 if the characteristic 
angular scale of variance $\theta_{disk}$ of the disk emission 
is larger than 
the angular width, $\theta_{FoV}$, of the field of view of 
telescope and the angular size 
 $\theta_{source}$ of gamma-ray source,   
method (a) works to overestimate 
 the flux     
 by an amount of $\sim \pi (\theta_{source})^2 \cdot F_{disk}$.  
In this context, method (b) can
 be trusted to yield the $\lq\lq$correct" flux, 
though it tends to be affected by neighboring sources to give a lower flux. 

When the wobble and ring background method (b) are utilized, 
the size of the gamma-ray emission, $\theta_{source}$, is presumed 
to be less than the radius of the ring. 
The condition     
for method (b) being useful can be expressed as  
 $\delta\theta_0 \ll \theta_{source} \ll \theta_{r1} \, 
({\rm or} \, \theta_{r2}) 
\approx \theta_{FoV}$. 
It might be said, in this context, that 
the angular resolution and the field of view of our observation is not 
very appropriate, namely, not good or wide enough,  to clarify    
the morphological structure peculiar to   
 the vicinity of PSR~B1706$-$44 by using method (b).       
 
In principle, and also in actual cases, it is neither easy nor clear  
to strictly distinguish the emissions of the Galactic disk and 
the individual objects from each other.   
The emission from the Galactic disk is likely to have a $\lq\lq$granular structure", 
which is caused by: the spatial distribution 
of energetic electrons with short lifetime by cooling, 
an irregular magnetic field of varying strength,  
 complex distribution of the matter density of molecular clouds, 
the diffusion process of cosmic rays 
escaping from the acceleration site, and so on. 
Thus, the structure of the disk emission of gamma-rays is considered  
to have a variety of spatial scales, varying from place to place,  
depending on the local environmental conditions of the Galactic disk. 

The $\lq\lq$granular structure"  
of the disk emission possibly lead us to 
erroneous conclusions,   
%failure in detection,
such as confusion concerning the source position,   
 incorrect estimation of angular size of the extended sources, or  
spurious detection of point-like sources etc., 
which are influenced also 
by the performance of the telescope in use and, in particular,  
by the scale size of the $\lq\lq$granularity" $\theta_{disk}$. 
The energy spectrum of the obtained gamma-ray flux can also be affected  
by the gamma-rays in the background data. 

%The bright optical lights along the Galactic plane can deform 
% the image parameters of IACT,   
%affecting an estimate of, for example, energy of TeV gamma ray.       
%The systematic error caused in the flux of TeV gamma rays      
%is estimated to be as large as ???\% 
%from applying several different ways of software padding to OFF-source data.    
  
\section{Conclusion}
  
Observations for about 50 hours with CANGAROO-III telescope system  
have given an indication of extended emission of TeV gamma rays 
 around the pulsar PSR~B1706-44.   
The strength of the signal depends on how we estimate  
angular size of the extended emission.  
%When the data was analyzed by using method (a) of 
%comparing the background events 
%with those in the region off the Galactic plane,    
The total flux at 1 TeV is 
$(4.7\pm 0.7)\times 10^{-11} (E/1{\rm TeV})^{-3.1\pm 0.7}$
cm$^{-2}$s$^{-1}$TeV$^{-1}$,    
when integrated for incident angles within a circle 
 of $1^o$ radius.         
This corresponds to    
$(4.9\pm 0.7)\times 10^{-8} (E/1{\rm TeV})^{-3.1\pm 0.7}$
cm$^{-2}$s$^{-1}$TeV$^{-1}$sr$^{-1}$ in unit of 
$\lq\lq$per solid angle".      
After  
integration of the gamma-ray energy, $E$,   
it is  2.2$\times$ 10$^{-11}$ cm$^{-2}$s$^{-1}$ for $E>$1TeV,    
which is as large as the Crab flux of 1.8 
$\times$ 10$^{-11}$ cm$^{-2}$s$^{-1}$. 

The intensity within the area corresponding to the point spread function,  
 $\theta<0.24^o$  from PSR~B1706$-$44,     
is $(3.0\pm 0.6)\times 10^{-12}$ cm$^{-2}$s$^{-1}$
for gamma-ray energy $E>$ 1 TeV.   
The flux corresponds to $17$\% of the Crab flux at 1 TeV,  
setting constraint on the emission from a compact source,  
which may underlie below the extended emission.    
The relative excess of this region compared with that of
$0.4^o<\theta <0.6^o$ is $6\pm 4$ \% Crab.

On the other hand,  
a statistically significant result  
could not be obtained 
from method (b) of the wobble and ring background analysis. 
The  2$\sigma$ upper limit 
on the emission within $0.24^o$ radius from PSR~B1706$-$44  
is $1.8\times 10^{-12}$ cm$^{-2}$s$^{-1}$ at 1 TeV, 
which corresponds to 10\% of the Crab flux at 1 TeV. 

The extended emission with method (a),  
together with the result using method (b),    
suggests  
complex structure of TeV gamma-ray emission 
existing in the vicinity of PSR~B1706$-$44.      
A deeper investigation and further efforts for improving the technique of 
Imaging Air Cherenkov telescope remain to be pursued,   
 in order to    
 resolve and distinguish Galactic sources of TeV gamma-rays    
against the diffusive emission of the Galactic disk.

\acknowledgments

This work is supported by a Grant-in-Aid for Scientific Research by
the Japan Ministry of Education, Culture, Sports, Science and Technology, 
the Australian Research Council, JSPS Research Fellowships,
and Inter-University Researches Program 
by the Institute for Cosmic Ray Research.
We thank the Defense Support Center Woomera and BAE Systems.

\end{document}